\documentclass[12pt,a4paper]{article}

\usepackage[utf8]{inputenc}
\usepackage{graphicx}
\usepackage{xcolor}
\usepackage{amsfonts}

\usepackage{amsmath,amssymb,amsfonts}
\usepackage{algorithmic}
\usepackage{subcaption}
\usepackage{graphicx}
\usepackage{textcomp}
\usepackage{xcolor}
\def\BibTeX{{\rm B\kern-.05em{\sc i\kern-.025em b}\kern-.08em
    T\kern-.1667em\lower.7ex\hbox{E}\kern-.125emX}}

\newcommand{\E}{\mathbb{E}}
\newcommand{\Prob}{\mathbb{P}}

\newcommand{\I}{\mathbb{I}}

\begin{document}

\title{Resource Adequacy and Capacity Procurement: Metrics and Decision Support Analysis}

\author{Chris J. Dent$^1$, Nestor Sanchez$^1$, Aditi Shevni$^2$, \\Jim Q. Smith$^2$, Amy L. Wilson$^1$ and Xuewen Yu$^3$ \\ \\
$^1$School of Mathematics, University of Edinburgh, UK\\ 
$^2$Department of Statistics, University of Warwick, UK\\
$^3$MRC Biostatistics Unit, University of Cambridge, UK\\
Corresponding author: chris.dent@ed.ac.uk}

\date{~}

%\keywords{resource adequacy, risk aversion, decision analysis, decision support, CVaR, wind power}

\maketitle

\begin{abstract}
Resource adequacy studies typically use standard  metrics such as Loss of Load Expectation and Expected Energy Unserved to quantify the risk of supply shortfalls. This paper critiques  present approaches to adequacy assessment and capacity procurement in terms of their relevance to decision maker interests, before demonstrating alternatives including risk-averse metrics and visualisations of wider risk profile. This is illustrated with results for a Great Britain example, in which the risk profile varies substantially with the installed capacity of wind generation. This paper goes beyond previous literature through its critical discussion of how current practices reflect decision maker interests; and how decision making can be improved using a broader range of outputs available from standard  models.
\end{abstract}

\section{Introduction}

%\bstctlcite{IEEEexample:BSTcontrol}

%\cjd{Distinctive feature is setting idea of prob dists in context of wider thinking about decision analysis -- others have .}

Power system resource adequacy (RA) is the field of managing the risk that there will be insufficient supply resource to meet electricity demand. Studies vary as to the precise class of events in scope, but is generally taken to encompass the balance of resources without considering fine detail of system operation (see \cite{stenclik} for a survey of current issues). This remains a topic of great interest, due to the need to maintain an appropriate level of system reliability as the profile of resources evolves towards a lower carbon portfolio -- and hence it is a key topic in assessing the performance of energy technologies within the system, and how technologies relevant to the net zero transition  complement each other  (or not) at whole power system level.

The area of RA also provides an excellent example of wider issues in project and policy appraisal, and in decision analysis, in that relevant decisions typically involves balancing capital investment costs (which are relatively concrete) against future reliability of the system (quantification of which is both more uncertain, and is not a cashflow that can immediately be incorporated in a monetised const-benefit analysis).

RA studies typically use a standard set of risk metrics to quantify system reliability: Loss of Load Expectation (LOLE), the expected (in the statistical sense) duration of shortfall in the future year or season under study; Expected Energy Unserved (EEU), the expected volume of energy demand not supplied; or if going beyond these, further summary statistics such as System Average Interruption Duration/Frequency Index (SAIDI/SAIFI) \cite{keane2011}. Formal decision analysis for capacity procurement  tends involve either a risk level target set with respect to one of these statistics, or a Cost-Benefit Analysis (CBA) in terms of a per-MWh Value of Lost Load (VOLL) multiplied by  EEU \cite{ZDWstorage,acer} -- historically less formally justified standards in terms of a deterministic measure of the margin of installed capacity over peak demand were common \cite{stenclik}, though these are now less widespread due to increased use of probabilistic analysis and the difficulty in including new technologies such as renewables without this.
%\cjd{Refer to scope as being options for utility functions used, rather than statistical estimation.}

There are issues in principle with the use of such summary statistics, which can have substantial practical consequences.
\begin{itemize}
    \item Decision makers are likely to be risk averse, and interested in variability of outcome in individual years, as well as an  average over possible outcomes. Expected value indices such as LOLE and EEU by definition do not reflect this, and thus decisions based on current decision analysis approaches might not properly reflect the concerns of decision makers. %\item There are also  less tangible issues such as wider economic confidence that are as important as direct customer effects. Thus conventional CBAs may take too narrow a view.
    \item One single-number index cannot capture the whole risk profile of a system, and if the mix of supply and demand changes there may be changes in risk profile that these indices do not capture. There is nothing fundamentally wrong with the use of summary statistics, but they must be customised to and reflect the needs of decision makers.
      \item Standard approaches do not specify who the decision maker is. For instance, stakeholders (including ultimate decision makers in governments) tend to be very risk averse about electricity security of supply, due to the consequences for society and the economy if confidence in the  electricity system decreases.
      \item The  disruption costs of events are often assessed on a narrow basis of direct costs to customers, and  not considering  points such as wider  societal and economic confidence in a robust electricity supply. %\footnote{Even within this `direct costs' picture, the view taken is typically quite narrow, assuming a uniform fixed cost per unit of energy demand not supplied.}
      \item It may not be natural to trade off investment and disconnection costs  on the same (usually monetary) scale, i.e. they may not be \emph{commensurate} \cite{selfnonsense} (the formal decision analysis term for whether quantities can naturally be compared on the same numerical scale). For instance, many customers are unlikely to be indifferent between having supply and being paid monetary compensation, though this is implicitly assumed in many analyses.
\end{itemize} 

Other works have looked for instance at use of multiple metrics \cite{stenclik}, inclusion of risk-averse metrics within optimization problems \cite{psr}, and construction of probability distributions of \emph{ex post} outcome metrics (\cite{sheehy} and references therein, and the Belgian standard references within the international comparison at \cite{burkeinternational}). This paper, however, goes beyond previous literature through its critical discussion of how current practices  reflect decision maker interests; and how decision making can be improved using a wider range of outputs available from standard risk model structures. For a general reference on relevant principles of decision analysis on which this work is based, including risk aversion, see \cite{smithbda}. 

This paper will first describe and critique present approaches to adequacy assessment and capacity procurement in Section 2, before presenting in Section 3 alternatives such as risk-averse metrics and broader visualisations of risk profile to support decision makers. This is illustrated with  results for Great Britain, considering a range of wind generation capacities.  Section 4 provides an extended discussion of issues of application and relevance to wider decision appraisal, and finally Section 5 presents summary conclusions.

%%%%%%%%%%%%%%%%%%%%%%%%%%%%%%%%%%%%%%%%%%%%%%%%%%%%%%%%%%%%%%%%%%%%%%%%%%%%%%%%%%%%%%%%%%%%%%%%%%
%%%%%%%%%%%%%%%%%%%%%%%%%%%%%%%%%%%%%%%%%%%%%%%%%%%%%%%%%%%%%%%%%%%%%%%%%%%%%%%%%%%%%%%%%%%%%%%%%%

\section{Standard picture of decision analysis for capacity procurement}\label{sec:standard}

%%%%%%%%%%%%%%%%%%%%%%%%%%%%%%%%%%%%%%%%%%%%%%%%%%%%%%%%%%%%%%%%%%%%%%%%%%%%%%%%%%%%%%%%%%%%%%%%%%

\subsection{Risk modelling for resource adequacy}

This section provides a brief overview of the resource adequacy modelling on which results presented, based on a GB system supplied by wind and conventional generation. This is satisfactory for our purpose of demonstrating use of model outputs in supporting decision making, where the key point is how a changing penetration of wind energy changes the overall risk profile. Issues of extension to other resources such as storage and interconnection to other systems, and uncertainty management arising from limited data on extremes and complexity of systems spanning multiple countries, will be discussed in Section \ref{sec:disc}.

We denote random variables with uppercase and constants with lowercase. $X_t, Y_t$ and $D_t$ denote available conventional and renewable generation, and demand, respectively at time $t$ in the future period under study; then the surplus $Z_t=X_t+Y_t-D_t$. This section is consistent with standard references such as  \cite{keane2011, singh}, but expressed in slightly different notation.
%; finally, we use subscripts to denote individual vector components, and superscripts to denote vector time indices when necessary.
% To avoid confusion with time indices, we denote vector components as superindices when necessary, so $\bb{y}_t^{(j)}$ is the j-th component of vector $\bb{y}$ at time $t$. Lastly, vector ratios and other opearators are applied componentwise.

\subsubsection{Non-sequential model}
\label{sec:single-area-system}
Let the period for which an assessment is being made (e.g. a future peak season) be divided into $n$ hours. LOLE, the expected number of hourly shortfalls in the period is\footnote{For simplicity, an hourly time step is used, as per available GB data -- in N America this `hourly' LOLE is usually called LOLH. We will also use data from the GB peak winter (Nov-Mar) season  -- as daily peak demands in winter are much higher than at other times of year, and  very low renewable output can occur at times of very high demand, the peak demand season dominates annual shortfall risk in GB. This observation would not apply if a large renewable capacity  shifts the time of year at which the highest values of (demand - renewables) occur -- however, as this paper is primarily about the choice of model \emph{outputs},  its conclusions are also broadly relevant to such systems.}: 
\begin{equation}
    \mathrm{LOLE} = \E\left[\sum_{t=1}^n \I(Z_t < 0)\right] = \sum_{t=1}^n \Prob(Z_t<0),
\end{equation}
and the EEU is defined as: 
\begin{equation}
    \mathrm{EEU} = \E\left[\sum_{t=1}^n \max\{0, -Z_t\}\right].
\end{equation}
This is commonly referred to as a non-sequential model, as unless there are technologies such as storage which explicitly link the system states at different times, the terms in the sum may be evaluated separately, and the times re-ordered without affecting the result.

For statistical modelling, it is often more convenient to work in a \emph{time-collapsed} picture with the time-collapsed variable $Z$ representing surplus at a randomly chosen point in time. For a system that does not have storage or other technologies which link time periods, the LOLE is then specified as 
\begin{equation}
    \mathrm{LOLE} = n \Prob(Z < 0) = n \Prob(X < D'),
\end{equation}
and an analogous formula applies for EEU. %\nsfixed{Because LOLE and EEU are aggregated over the entire time period, these risk metrics only depend on the marginal distributions of surpluses, hence a time-collapsed approach is suitable.}

The most common means of estimating the distribution of $D_t-Y_t$ is to use the empirical historic data directly in predictive risk calculations, sometimes referred to as  \textit{hindcast} \cite{wilson2019using, keane2011}
%The distribution of net demand is then given by
%\begin{equation}
%    \label{eq:hindcast}
%    \mathbb{P}(D - Y \leq w) = \frac{1}{T} \sum_{\tau=1}^T \I(d_\tau - y_\tau \leq w).
%\end{equation}
%where $T$ is the number of observations in the historic record,  $\tau$ indexes the historic records, and historic demand and wind resource are rescaled appropriately to the future system scenario under study. This approach may also be interpreted as estimating the risk conditional on a repeat of historic conditions in one or more years.
The hindcast estimate of LOLE conditional on a particular historic weather year $y$ is then
\begin{equation}
    \label{eq:hindcastlole}
    \mathrm{LOLE}_y = \sum_{\tau \in T_y} \Prob(X_{\tau} < d_{\tau}-y_{\tau})
\end{equation}
where $T_y$ is the set of times in historic year $y$, $d_{\tau}-y_{\tau}$ is written in lower case as it is historic data rather than a random variable, and \emph{historic} times are indexed by $\tau$. LOLE conditional on no particular weather year is then usually estimated as $\mathrm{LOLE} = (1/n_Y)\sum_y \mathrm{LOLE}_y$, where $n_Y$ is the number of years of data (and similar for EEU). As discussed later, there can be substantial uncertainty in any estimate of unconditional risk level, as the estimae of the mean will often be dominated by a small proportion of the historic years.

\subsubsection{Time sequential model}

For   model outputs beyond the standard expected value indices, such as the distribution of energy unserved, or the distribution of Loss of Load Duration (LOLD, the random variable of which LOLE is the mean) a time sequential model would be required. LOLD is specified as %$\mathrm{LOLD} = \sum_{t=1}^n \I(Z_t < 0)$
\begin{equation}
    \mathrm{LOLD} = \sum_{t=1}^n \I(Z_t < 0),
\end{equation}
and the  energy unserved (EU) as %$\mathrm{EU} = \sum_{t=1}^n \max\{0, -Z_t\}$.
\begin{equation}
    \mathrm{EU} = \sum_{t=1}^n \max\{0, -Z_t\}.
\end{equation}
A wide range of other possible outputs may also be calculated, the usual mechanism for doing so being Monte Carlo simulation.

Stochastic process models must then be estimated for $X_t$ and $(D_t,Y_t)$. In practice, again a common way to proceed is to use a hindcast estimate for the process of demand and wind, i.e. %$\mathrm{LOLD}_y = \sum_{\tau \in T_y} \I(X_{\tau} < d_{\tau}-y_{\tau})$, 
\begin{equation}
    \label{eq:hindcastlold}
    \mathrm{LOLD}_y = \sum_{\tau \in T_y} \I(X_{\tau} < d_{\tau}-y_{\tau})
\end{equation}
and similar expression  for EU. The aggregate available conventional capacity is usually specified as a sum of stochastic process models for each individual unit, with the unit models commonly (as here) being two state Markov birth-death processes.

%%%%%%%%%%%%%%%%%%%%%%%%%%%%%%%%%%%%%%%%%%%%%%%%%%%%%%%%%%%%%%%%%%%%%%%%%%%%%%%%%%%%%%%%%%%%%%%%%%

\subsection{Decision analysis for capacity procurement}

It is common to look at capacity procurement for a single future year, which for simplicity we will do here. This is the approach taken in the GB capacity market, where  a target risk level is set based on a cost-benefit analysis (CBA) for the future year considered \cite{GBcm}, or might represent a long run equilibrium problem \cite{bothwellej}.
%(in GB the main auction is held 4 years ahead of the delivery year, with an additional auction held 1 year ahead) 
 
%\footnote{It is not guaranteed that procuring capacity looking at just one target year will bring forward an ideal portfolio looking over multiple future years.} 
% \cjd{Mention in passing that long run eq never occurs, and that looking myopically at one year may not bring forward a good profile of capacity in the longer term inc regulatory risk.}

The standard CBA is then be expressed as an optimization problem \cite{ZDWstorage}:
\begin{equation} \label{eq:cba}
    \mathrm{min} ~~~~~~ c(R) + [\mathrm{VOLL}] \times [\mathrm{EEU}](R)
\end{equation}
over the possible sets $R$ of capacity-providing resources \cite{ZDWstorage}. Procurement cost $c$ and EEU are both functions of $R$. This is commonly simplified \cite{acer} to 
\begin{equation} \label{eq_conevoll}
    \mathrm{min}~~~~~~[\mathrm{CONE}]\times r + [\mathrm{VOLL}] \times [\mathrm{EEU}]_r,
\end{equation}
on the assumption that the additional procured capacity simply shifts the probability distribution of surplus/deficit by the mean available capacity $r$ from the addition\footnote{This is justified either through convolution of an independent addition with a distribution of deficit that is approximately exponential in shape in the relevant region \cite{chris-stan}, or through the Central Limit Theorem if a small independent addition is made to a large background of independent units.}. In practice, this assumption  implies that the addition should be small compared to the resource already present (generally the case in capacity markets, where the volume of new capacity is usually limited); and does not contain renewable generation or storage for which independence between existing and additional units does not apply. 
Here Cost of New Entry (CONE) and VOLL take fixed values, and $[\mathrm{EEU}]_r$ is the expected energy unserved if the volume of capacity procured is $r$; it is straightforward to generalise this to non-constant CONE and VOLL.

\subsection{Consequences of choice of VOLL}

At the value of $r$ which minimises (\ref{eq_conevoll}), 
\begin{equation}
    [\mathrm{LOLE}]_r=\frac{[\mathrm{CONE}]}{[\mathrm{VOLL}]}
\end{equation}
There are various challenges to taking this as an `optimal' solution for the real world, even if  EEU is regarded as a sufficient summary statistic of risk profile.
\begin{itemize}
    \item Studies usually implicitly assume that the control room can disconnect  the amount of load required and no more, with perfect foresight. This is not the case in practice, which implies  an  increase in energy unserved.
    \item An average VOLL across all customers is usually used, i.e. the interests of customers who are relatively indifferent to disconnection are treated interchangeably with interests of customers who are more averse to disconnection, even if there is no discrimination as to who is disconnected involuntarily. If one accepts the idea of monetising unserved energy using VOLL, should customers' interests be averaged in this way, or should more weight placed on the interests of customers who are more inconvenienced by being disconnected?
    \item As  in the introduction, one needs to consider whether reliability and procurement costs are commensurate, i.e. whether they can be compared on the same numerical scale.  
\end{itemize}
%Even within the standard  picture of (\ref{eq_conevoll}), making different judgments associated with the first two of these points could change the `optimal' level of reliability very substantially. 

This standard optimization picture in (\ref{eq_conevoll}) typically recommends a level of reliability similar to the GB standard of 3 hours/year LOLE, whereas a system that unreliable would probably be deemed politically unacceptable. For instance, in GB, a system margin warning can be a major news story \cite{nism}, whereas this lowest level of system warning actually means that some hours ahead of real time the operator was not certain of having their usual real time operating headroom, i.e. nowhere near a shortfall in real time -- if actual real time shortfalls happened in a substantial proportion of colder winters, the reaction in public debate would be  stronger still. 

Using a higher (not average) VOLL, based on the second bullet above, would push the reliability standard more towards a level that  would be considered  acceptable in this wider sense but would not consider wider issues of societal confidence in the electricity supply. Indeed if such factors beyond individual customer damage are considered important, then for use in comparing different capacity portfolios VOLL might be chosen such that (\ref{eq_conevoll}) gives an acceptable level of system-level reliability, rather than being based on customer surveys. It is certainly the case that making different judgments associated with the first two bullets can change the `optimal' level of reliability from (\ref{eq_conevoll}) very substantially. 

%\cjd{Mention somewhere the different between society averaging and individuals experiencing events rarely, how this affects risk aversion, perception of different classes of disconnection event.} 
%These two sensitivities alone could decrease the `economically optimal' reliability standard by a factor of 5 or more, depending on the judgments taken on these two bullet points.

%It seems natural to hypothesise that these two sensitivities could alone shift the `economically optimal' reliability standard by a factor of 5-10, depending on what judgments are taken -- and this is before any consideration of uncertainties arising from statistical estimation or model structure discrepancy.  

\section{Beyond the conventional framework}\label{sec:beyond}

\subsection{Background and multicriteria formulation}

%\subsubsection{Background}

Most formal decision analysis frameworks for capacity procurement assume the approach described in the previous section, i.e. monetising future reliability in terms of VOLL multiplied by expected energy unserved. 
Clearly capital costs are naturally in terms of money, though there may be uncertainty in the monetary sum, or one may wish to use a capacity price curve (i.e. as more is procured, the unit cost of capacity increases) rather than a fixed CONE -- we have already discussed whether this can naturally be compared with reliability on the same numerical scale.
% One may also need to consider questions of cost `to whom', which again may link back to how broadly the decision question is posed (the narrom question of capacity procurement for a single year, wider issues of how the capacity will generate energy over multiple years, etc.)
%The `reliability cost' term does not, however, naturally come as a monetary sum, and we have already discussed above the nuances associated with monetising such  matters of  social value. 
Moreover, expected monetary return is rarely a utility function that reflects decision makers' interests, particularly for mitigation of rare high impact  events, and thus introducing a degree of risk aversion seems natural.  

The simplest evolution of (\ref{eq_conevoll}) would be to extend this framework to a multi-criteria decision question, seeking a Pareto frontier on which EEU cannot improved without disbenefit in terms of cost, and vice versa. However, for this  case mapping the Pareto frontier is in fact equivalent to a sensitivity analysis on VOLL, so results for it are not presented here.  %\cjd{(note that one difficulty bringing very sever outcomes into the formal analysis is that we might not have the data to quantify their prevalence by bottom-up probability modelling, see `other considerations'.)}

%\subsubsection{Example - expected value indices}

%\cjd{Make consistent with Nestor's notation} If one minimises (\ref{eq_conevoll}) for fixed CONE and VOLL, the resulting system has an LOLE equal to CONE/VOLL. One could alternatively regard this as a multicriteria optimization minimzing 
%\begin{equation} \label{eq_multcrit}
%    \lambda\times x + [\mathrm{EEU}]_x,
%\end{equation}
%where the weighting $\lambda$ is the LOLE level at the optimum. it is then, for instance, straightforward to plot a Pareto frontier in the space of $(x,[\mathrm{EEU}]_x)$ showing the solutions for which reliability cannot be . \cjd{Need to do this for our example, with appropriate axes of capital cost and either EEU or reliability cost.}

\subsection{Data for examples}\label{sec:data}

The following sections include results from an exemplar based on the Great Britain (GB) system. The standard approaches described previously are used for risk calculations. A comparison between scenarios of different installed wind capacities is made by using the same portfolio of available conventional capacity for each scenario; and for each scenario of installed wind capacity shifting the supply-demand balance to give a common EEU of 3 GWh/year. For instance, one might hypothesise that at higher penetrations of renewable generation, for a given value of standard indices such as LOLE or EEU, greater variability of supply  leads to greater variability of outcome. This comparison is controlled in the sense that it looks at overall risk profile for a range of scenarios which have the same value of the headline EEU.

The scenario of installed conventional capacity is based on one originally provided by National Grid ESO, with a small random element added to each capacity due to the commercial sensitivity of the raw data. For sequential models, availability probabilities from NGESO are supplemented with mean repair time data from the IEEE Reliability Test
System \cite{ieee-mttr}.

GB demand data for the 12 peak (winter) seasons 2005-17 are used, with an estimate of historic available embedded renewable capacity added back on. Demand data are rescaled to a common system scenario according to historic values of the Average Cold Spell (ACS) Peak statistic, with the given scenario defined by an ACS value.

The wind generation data used are from \cite{renewables-ninja}, and combine historic reanalysis wind speed data with a future scenario of what is connected to the system. Hourly capacity factors (CFs) for onshore and offshore wind in the  `near term' wind fleet from \cite{renewables-ninja} are used, and for a given scenario of onshore and offshore wind capacity connected to the system these hourly CFs are multiplied by the respective GW installed capacities and added to give a total hourly available capacity. The division between onshore and offshore for a given level of total installed wind capacity is based on projections in \cite{nestorthesis}.

Thus our exemplar system is generally representative of the GB system and will suffice for our illustrative purpose, however for a fully applied GB study one would need to use data that are  specialised to the particular decision question considered.

\subsection{Risk averse metrics}

\subsubsection{Background}

It is possible to define alternative single number metrics which give a measure of risk aversion, for example the well known (Conditional) Value at Risk (VaR and CVaR) \cite{cvar},  with respect to model outputs such as the constructed distribution of energy unserved. This is defined as for a random ariable $U$ as 
\begin{equation}
    [\mathrm{CVaR}]_{\alpha}=\E[U|U>u']
\end{equation}
 where $P(U\ge u')=1-\alpha$ for a risk threshold parameter $alpha$. Thus the mean of $U$ is the special case $[\mathrm{CVaR}]_0$, and if $U$ is the  energy unserved then CVaR is a generalisation of the standard EEU index -- while one can use other mappings of EU and LOLD to give risk averse utility functions, this would not have the same attractive property of  generalising the expected value indices.% (this is in addition to CVaR's attractive general properties within optimization problems.) One could also use other mappings of the distribution of e.g. EU or LOLD as alternative risk averse utility functions.

CVaR has the further beneficial property of convexity when embedded in a wide range of optimization problems \cite{cvar}. We do note however that where it is not necessary to embed within an optimization model, there can be difficulties in communicating CVaR results outside the specialist community, particularly as for low degrees of risk aversion the risk threshold $u'$ will sit in the other tail of the distribution of $U$ from the one of interest. Another diadvantage is that CVaR values with different $\alpha$ are not directly comparable, despite having the same dimensions.

Calculating risk-averse indices will in general require time-sequential modelling. The exception would be to work with VaR and CVaR with respect to the snapshot LOLP or Expected Power Unserved (i.e. the probability of a shortfall, or expected shortfall, at a randomly chosen point in time \cite{chris-stan}). This would, however, need to be interpreted carefully. For instance, VAR with respect to snapshot LOLP may have a useful interpretation in terms of the expected number of hours with surplus below a given level,  but other combinations of VAR/CVaR with LOLP or EPU might not be so interpretable.

%One can produce various representations of how risk aversion affects conclusions of studies, for instance looking at CVaR as a function of $\alpha$ for different scenarios; calculating capacity value metrics with respect to risk averse metrics; and using a risk averse metric in a multicriteria  analysis for capacity procurement. However, on the latter option, there is an issue in principle of CVaR values for different $\alpha$ not being directly comparable, despite having the same dimensions. 

\subsubsection{Example}

\begin{figure}[h!]
    \centering
    \includegraphics[width=0.7\textwidth]{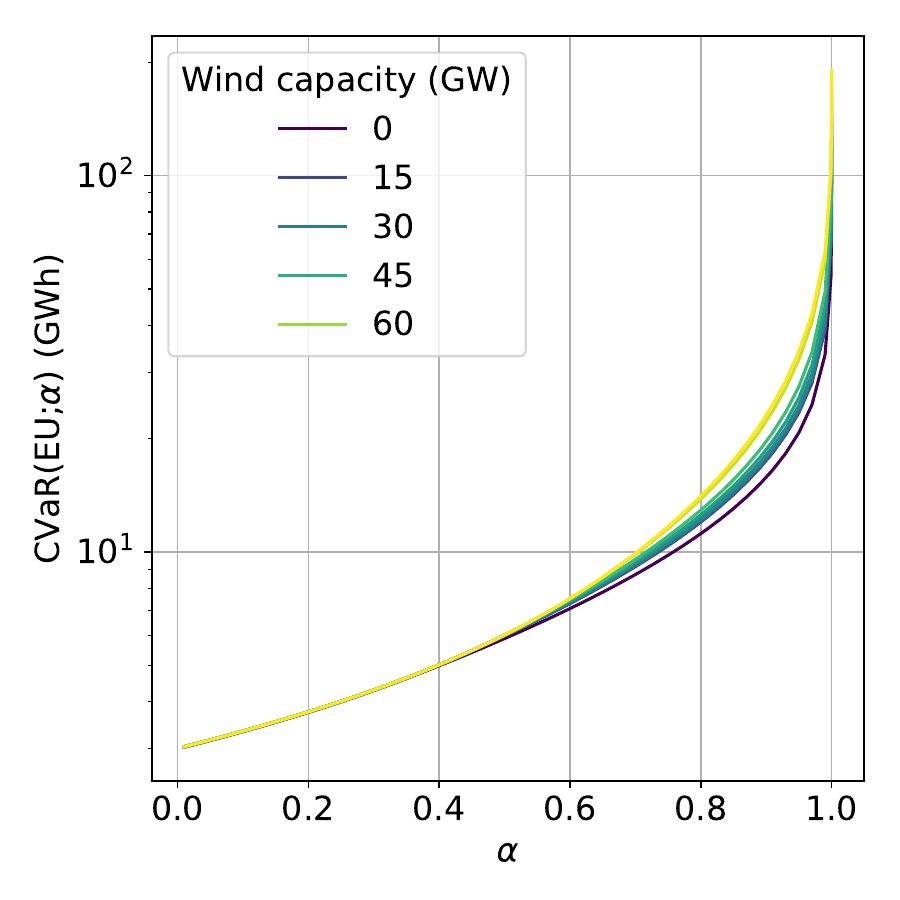}
    \caption{CVaR as a function of the risk aversion parameter $\alpha$ for a range of installed wind capacities.}
    \label{fig:ratio-vs-alpha}
\end{figure}

%\begin{figure}
%    \centering
%    \includegraphics[width=\columnwidth]{decision-analysis-gb.pdf}
%    \caption{Comparison of CVaR based results for a GB example, with a range of installed wind capacities}
%    \label{fig:GB_VaR}
%\end{figure}

Fig. \ref{fig:ratio-vs-alpha} shows CVaR with respect to EU as a function of $\alpha$, for a range of installed wind capacities. As in all examples, for a controlled experiment the scenarios of different installed wind capacity have the same EEU. As anticipated in the section on data, as the wind capacity increases, the CVaR for a given $\alpha$ also increases. This is consistent with a hypothesis that greater variability of supply would lead to greater variability of outcome -- however the effect is not very large, and the next section will explore how CVaR as a summary statistic does not reveal the most striking change in the risk profile at higher wind capacities.

\subsection{Visualisations for decision makers}

%\subsubsection{Background}

Instead of attempting to define metrics in this way, one could instead provide visualisations to decision makers of the consequences of particular decisions in different scenarios of planning background, and  let them decide on that basis how much capacity to procure. Even if there is still a preference of working with summary statistics for formal decision analysis, there is value in supplementing this with a wider range of visualisations to understand more broadly the system's risk profile, or the consequences of results from formal optimization models.

This is an attractive idea in principle, though to go with such visualisations one needs the necessary skills in how to use them well -- both on the part of the analysts in terms of how to create visualisations, and also in terms of how some technical statistical understanding on the part of decision makers may be necessary. On the other hand, visualisations such as these may be more tangible and easier to communicate than summary statistics such as CVaR. Appropriate use of scenarios and supporting narratives can help here, potentially using an interactive dashboard to provide model outputs and narratives of consequences for decisions under consideration.

\begin{figure}[h!]
    \centering
    \begin{subfigure}{0.45\textwidth}
    \includegraphics[width=\textwidth]{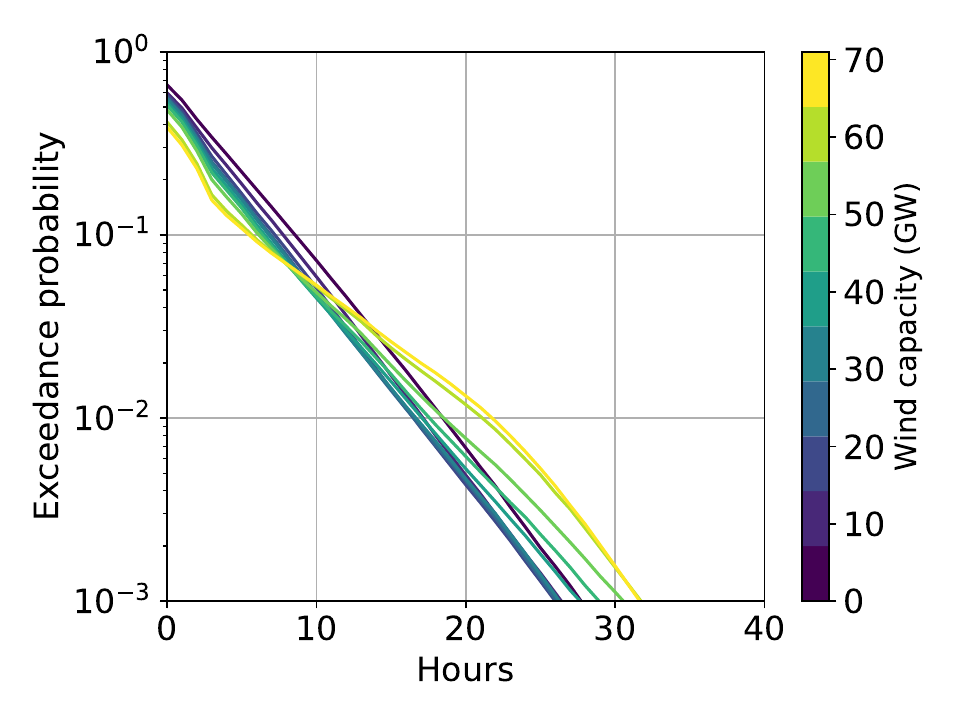}
    \caption{Loss of load duration}
    \end{subfigure}
    \begin{subfigure}{0.45\textwidth}
    \includegraphics[width=\textwidth]{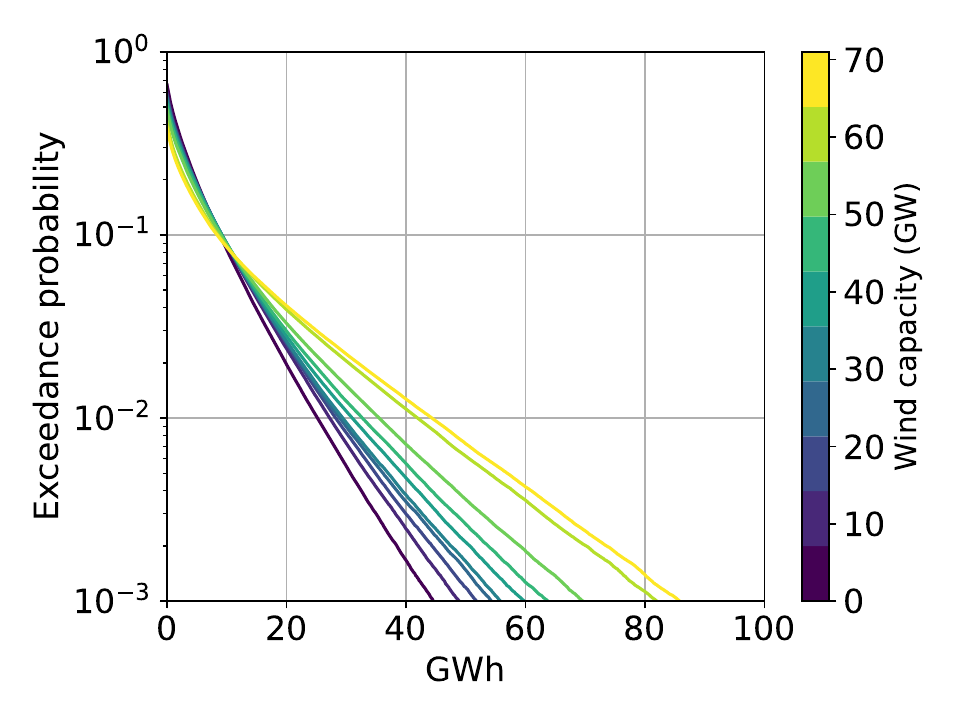}
    \caption{Energy unserved}
    \label{fig:with05}
    \end{subfigure}
    \begin{subfigure}{0.45\textwidth}
    \includegraphics[width=\textwidth]{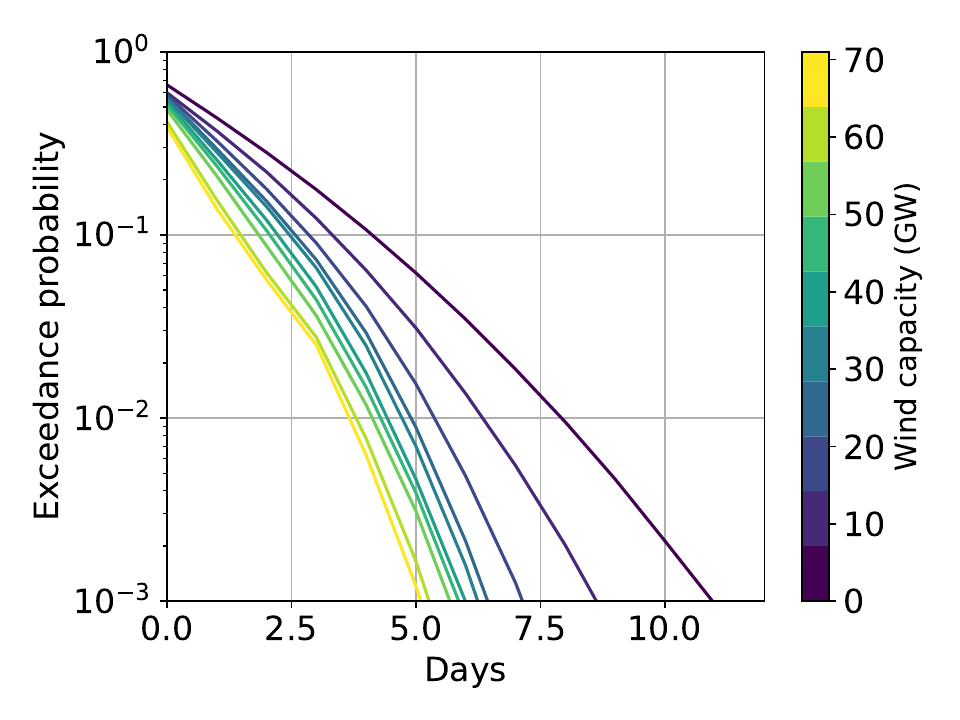}
    \caption{Number of days of shortfall}
    \end{subfigure}
    \begin{subfigure}{0.45\textwidth}
    \includegraphics[width=\textwidth]{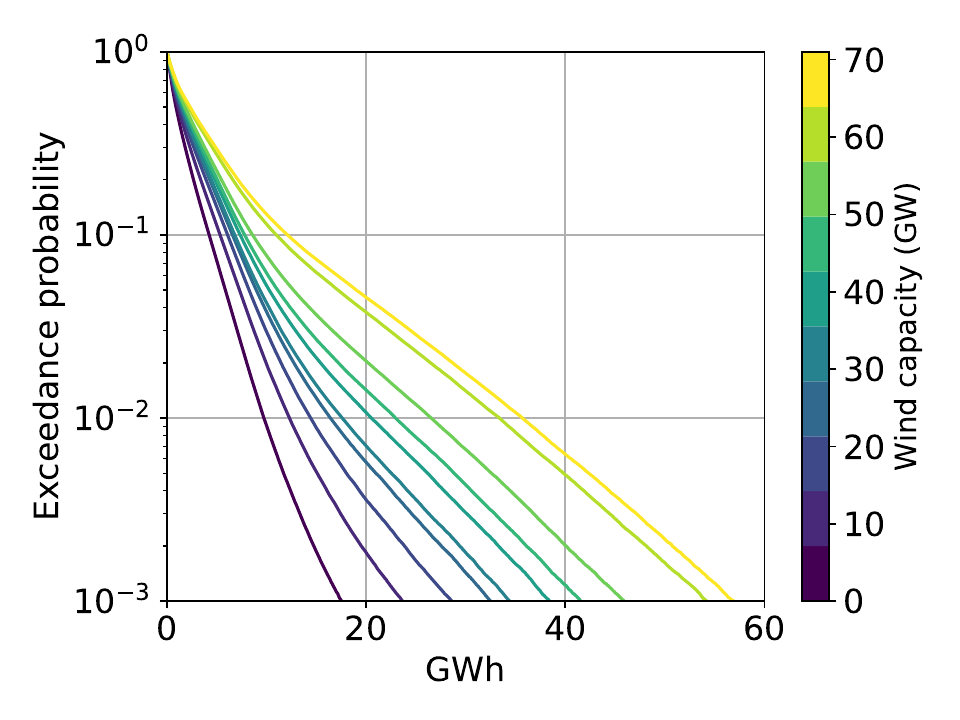}
    \caption{EU within days of shortfall}
    \end{subfigure}
    \caption{Scenario with fixed EEU = 3 GWh/year}
    \label{fig:fixed-eeu}
\end{figure}

%\begin{figure*}
%    \centering
%    \includegraphics[width=15cm]{GB-EEU=3318.png}
%    \caption{Results from sequential model with EEU fixed at 3.318 GWh/year.}
%    \label{fig:GB_viz_EEU}
%\end{figure*}

%\begin{figure}
%    \centering
%    \includegraphics[width=\columnwidth]{GB-LOLE=3.png}
%    \caption{Results from sequential model with LOLE fixed at 3 h/year.}
%    \label{fig:GB_viz_LOLE}
%\end{figure}

Fig. \ref{fig:fixed-eeu} provides an example of this for a scenario with  fixed EEU of 3 GWh/year, and for a range of installed wind capacities. There is substantial variation with installed wind capacity of the distribution of EU  in our experiment, but rather less variation in the distribution of LOLD. However while the former explains the variation of CVaR with respect to EU, the CVaR being a summary statistic of the distribution of EU, the difference between scenarios of different wind capacities is less striking in the CVaR results than in the underlying probability distribution. %\cjd{Nestor do we have the CVaR on a linear scale as well as a log scale?}

One can look further at the overall risk profile through the distributions of number of days with a shortfall, and of the probability distribution of EU within a day conditional on that day having a shortfall. It is clear from the lower two panels of Fig. \ref{fig:fixed-eeu} that at higher wind penetration the same EEU is made up (in a probabilistic sense) of fewer days of greater shortfall. The variation with wind capacity of these two distributions is rather greater than that of the distributions of LOLD and EU -- this difference may well be material for decision making, demonstrating how  one might go in to considerable detail of risk model outputs to understand fully how the risk profile is changing.

There are important caveats on these results as quantitative calculations for the real GB system. We have carried out a calculation using a standard approach to illustrate important issues for decision support analysis in the model-world, but for a fully applied study one would need to specialise the statistical estimation to the relevant scenario, and to consider whether the modelling assumptions made give meaningful results on low probability events with respect to the real world.

%One final note in interpreting the results of these plots -- I do not think we or anyone else would claim that outputs of this model are meaningful for 1 in 1000 year (or less likely) events due to both parametric and model structure uncertainty.

\section{Discussion}\label{sec:disc}

%\subsection{Use of metrics in resource adequacy}

\subsection{Statistical issues}

In addition to this paper's technical work exploring the range of model outputs available from standard calculations and how these might be used, there are also important statistical and uncertainty management issues which must be considered. These include estimation of model inputs, including having a limited number of examples in the historic record of extreme conditions,  estimates of the availability properties of conventional generation at these times, and modelling of very complex continental-scale systems. 

It is common in reliability analysis to have limited direct historic  data on failure events, particularly in an area such as resource adequacy where there is a specific external driver of stress events (i.e. the weather). The consequent uncertainty  in estimation of model outputs tends to increase in systems with high renewables penetrations, where the highest values of (demand minus renewables) are from times combining sufficiently high demand with  low renewable resource -- this tends to concentrate risk in a smaller number of events as compared to a system where risk is driven by the highest values of demand. 

Strikingly, these  conditions that drive risk at hig renewable penetrations would otherwise be regarded as extreme in any one location, where they would appear to be a standard and  benign winter day with fairly low temperature and little wind. This dominance of risk profile by a limited number of years will manifest itself in different ways depending on the way the demand and renewable resource interact, and how much storage is connected to the system; for an example with a high storage penetration see \cite{zacharyRS}. 

\begin{figure}[h!]
    \centering
    \includegraphics[width=0.7\textwidth]{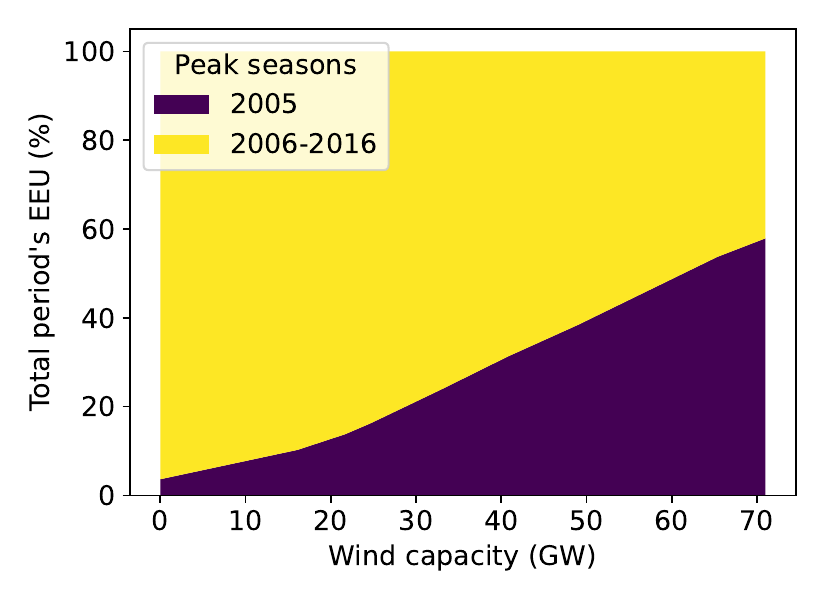}
    \caption{Proportion of estimated EEU arising from 2005 data.} 
    \label{fig:EEUvsWind}
\end{figure}

This effect is illustrated for the example of this paper in Fig.~\ref{fig:EEUvsWind}. At low penetrations of wind generation, 2005 does not make a substantial contribution to the calculated EEU, as peak demand was quite modest that winter. However, due to the very calm conditions on certain days of fairly high demand, at high wind penetrations this becomes the most significant winter by far in driving the outcome of the risk calculation.

\begin{figure}[h!]
    \centering
    \includegraphics[width=0.7\textwidth]{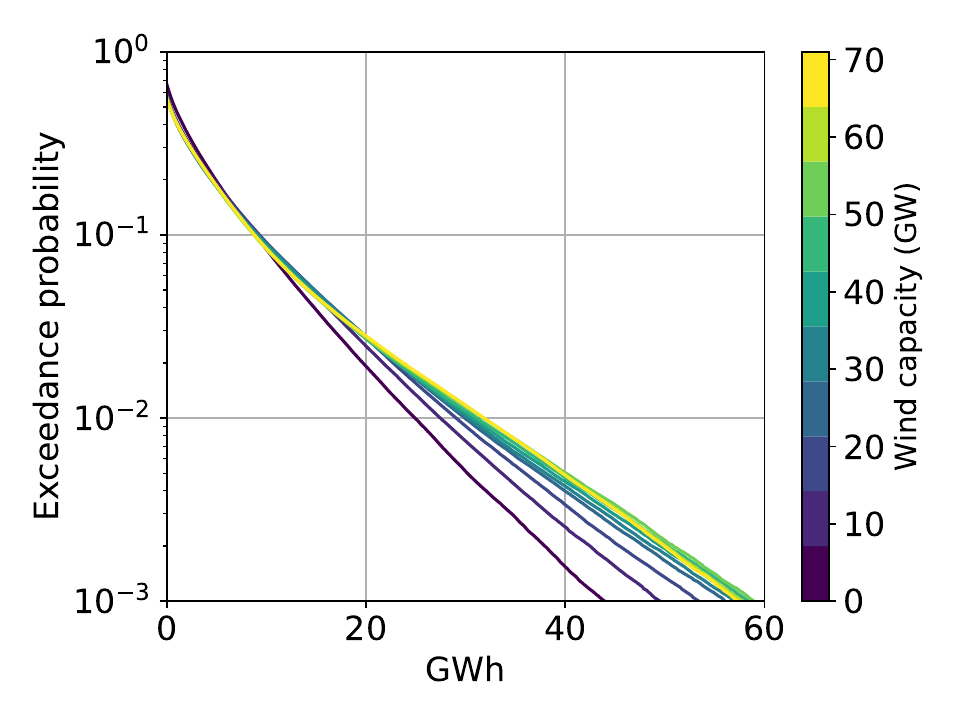}
    \caption{Probability distribution of Energy Unserved (same experiment as Fig. \ref{fig:with05} except that data from 2005-6 are omitted).} 
    \label{fig:no2005}
\end{figure}

As a further illustration based on the experiment in this paper, Fig.~\ref{fig:no2005} presents the same analysis as Fig. \ref{fig:with05}, except with the data from winter 2005-6 omitted\footnote{A comprehensive description of the experiment in this Figure is as follows. For each wind capacity, the data from all years are normalised to give an EEU of 3000 MWh/y in a calculation using data from 2006/7 onward, and then those same years of data from 2006/7 onward are used in the predictive risk calculation which outputs the distribution of EU.}. At high wind penetrations, the estimated probability of very severe outcomes is then much lower than if data from the outlier winter 05-06 are included. %Again, the origin of this is periods in 05-06 when temperatures were fairly low (and thus demand fairly high), but the wind conditions were very calm across the system. 

A particular manifestation of this will be in estimating tail event metrics such as VaR or CVaR of energy unserved with a high threshold - for instance in assessing what a 1 in $n$ year risk level might be. Even if one assumes that past weather is statistically representative of the relevant future year, by definition one will pick up a relevant weather year on average once every $n$ years. Historic data will thus be very sparse, and estimates of such metrics (not conditional on particular weather) will be speculative.

%\cjd{Decide whether to mention temperature dependence of mechanical availability in any detail, as it might be going a bit much into the weeds to have more than a sentence.}

Another key issue of uncertainty management is interconnection across very wide (continental scale) areas. As GB is an island system, this  manifests as undersea connections to Ireland, Norway and the main continental European system. The level of interconnection between GB and other systems is forecast to rise rapidly -- for instance one scenario study by the system operator has a minimum of 20 GW of connections to other systems on GB peak demand which is currently a little over 50 GW, making interconnector support  a key consideration in evaluating adequacy risk \cite{interconnectorNOA}. However there will be considerable uncertainty in statistical characterisation of available support from other systems, due to the volume of model assumptions and data involved in continental scale assessments, and when carrying out a study knowledge of other systems almost certainly being poorer than knowledge of one's own. The need to bring interconnection into assessments is recognised in other systems, as discussed in \cite{stenclik}.

\subsection{Decision support for capacity procurement}

There is broad recognition that planning for system adequacy needs to recognise the changing needs of present and future systems with high renewable penetrations \cite{stenclik}. In particular, it is necessary to recognise that if the profile of supply and demand changes in a system, then it may be difficult to track this using a small set of single-number metrics -- this paper has demonstrated by numerical experiment how this can be the case in a system with increasing penetration of renewable generation. Decision support approaches must both be transparent to the decision maker and not oversimplify the situation.

Historically, resource adequacy standards have tended to be set in terms of a single metric, usually a variant of LOLE -- either the expected number of days on which there is a shortfall (the classic LOLE standard in N America being 0.1 days/year LOLE \cite{billinton2014}), or the expected number of hours of shortfall (e.g. the GB standard of 3 hours/year LOLE; this metric is usually referred to in N America as Loss of Load Hours, LOLH, though in Europe is more commonly referred to as LOLE). 

There is thus a tension between the need to reflect the increasing complexity of patterns of supply and demand, and the natural desire to have simple, transparent criteria for use in supporting procurement decisions. The structure of working in this more complex environment would be simpler if there is a single `controlling mind' (i.e. a point of decision making in a single organisation, maybe with the decision vested in a single individual) which is able to take judgments as to how to balance different aspects of risk profile in a decision on capacity procurement. In that case, standard approaches to uncertainty management and multi-criteria decision analysis could then be used to handle the issues described. 

However when there is an industry and policy need for a clearly defined standard, one must seek appropriate compromises which reflect the changing nature of power systems and are sufficiently grounded in the relevant decision science; there is unlikely to be one definitive best approach here applicable in all situations. A further consideration is  what actually \emph{can} be estimated confidently -- in particular, in many systems it may be possible to make confident estimates of risk profile conditional on a particular weather year and on a statistical characterisation of available support from interconnectors, these being the principal sources of uncertainty in risk model outputs.

One basis for a solution could therefore be to set a public facing risk target, for use in capacity procurement, conditional on a single given weather and interconnector scenario, or an assumption that the available time series data  characterise fully the range of future possibilities that might be faced. This would allow performance of alternative resource portfolios to be compared, but it might be necessary to check against other possible interconnector or weather scenarios that the  outcome is likely to be robust. It may further be necessary to adjust either the risk target or scenario choice from time to time, in order to maintain an appropriate level of risk as the portfolio of technologies develops. A related challenge is that supply capacity is not a simple additive commodity of the kind that is traded in standard auctions; a fully discussion of the consequence of this may be found in \cite{ZDWstorage}.

It may also be necessary to make some special consideration of very extreme situations. If it is deemed infeasible to assign probabilities to events which are too sparse in the historic record, then these might be treated through scenario analysis -- however if this means that there is a probability model used for some classes of event, but the events which really matter are excluded from this model, there might be logical gap in the framework. 

Another situation might be where weather beyond a certain degree of severity introduces distinct failure modes which are not seen at all in less extreme, but still severe, conditions, for instance those of Texas in 2021 \cite{utexas}. A further example is the severe cold and snow in GB in 1962-62 \cite{rms}, which would bring very substantial changes in demand patterns over an extended period of months if it occurred again. Apart from any difficulties with assigning probabilities or mean return times, such conditions might require bespoke studies of effect of temperature and snowfall on demand and the power system, going well beyond the  modelling that suffices for more normal circumstances.

Finally we note that the transparency of metrics and visualisations for decision makers is a significant issue. Standard indices such as LOLE and EEU are sometimes regarded outside the specialist modelling community as as being hard to understand, and inside that specialist community there is recognition of limits on the information that they contain. It is likely that non-specialists will find some additional metrics such as CVaR more difficult still to understand, so it is vital to take appropriate care in communicating the information that they contain. It may be, however, that visualisations of probability distributions are easier to communicate in that they contain all the information on estimated statistics of a particular quantity, though initially some might find the  overt presentation of probability distributions intimidating.

\subsection{Relation to wider issues of project and policy appraisal}

The questions discussed in this paper should also be seen in the wider context of decision analysis and project appraisal. Indeed this area of resource adequacy and capacity procurement often provides a very good exemplar of wider issues, in that with a fairly simple system model (at least for a single power system area), it presents a range of subtle statistical issues.

The master guide for such analysis in the public and regulated sector in GB is the HM Treasury Green Book \cite{greenbook}; while this is explicitly about appraisal and evaluation of policies and projects in central government, it is very influential on decision analysis in wider contexts. One key principle is that categories of assessment in an appraisal should be monetised if possible; if monetisation is not possible they should be quantified on a different scale, and if quantification is not possible they should be assessed qualitatively.

There is no dispute that RA risk is subject to quantification, albeit with significant caveats on whether confident estimates can be made without conditioning on assumptions about weather and interconnector support. However, as mentioned earlier in this article, monetising RA risk is much more doubtful -- and the standard monetisation in terms of expected energy unserved multiplied by a (fixed, survey based, averaged over customers) per MWh value of lost load, tends to recommend unacceptably low level of reliability. In addition to this top-down argument, there are bottom-up arguments against averaging interests of all customers including that disconnections do not discriminate between customers with different interests.

One could go beyond the standard monetisation by making the VOLL dependent on the depth and length of shortfall, and more generally by eliciting decision maker judgments as to how the economic value should be considered; or by considering variability of economic damage about the mean. However, this still runs into problems with scope: does one only include direct economic damage? or does one wish to include in decision analysis wider issues such as the need for wider economic and societal confidence in a reliable electricity supply? The latter is likely to be of interest to high level decision makers.

This in turn is a wider example of the issue in appraisals of managing imprecisely defined concepts such as social value, environmental capital and cultural value; energy security may be regarded as an example of the former. There will usually then be uncertainty arising from questions of scope, and of the way that any monetisation is conceptualised. This means that uncertainty in monetisations goes beyond considerations of input, parameter and modelling uncertainty to  uncertainty in the conceptualisation (essentially that equally reasonable and expert people might come up with different broad approaches). The problems with bringing such disparate criteria together into a single number score has been noted previously in an energy context by Hammond and co-authors \cite{hammond2008,hammond2015}. A very prominent example in the wider GB infrastructure sector is the business case underpinning the present HS2 high speed rail project, in which many factors are combined into a single monetary CBA \cite{hs2updated}.

Going back to the specific case of resource adequacy, consequence is clearly quantifiable, and relative economic consequence between different scenarios is likely also be quantifiable -- but on the latter, questions of scope might mean that this is naturally a multicriteria comparison. The problem comes with the final step of bringing everything together into a single line item of money, when procurement costs and reliability are not fully commensurate. In this and a very wide range of case studies in other domains, we would recommend the following:
\begin{itemize}
    \item If monetisation  is used, it should come with a sufficiently broad quantification of uncertainty, including consequences of social value etc. being imprecisely defined quantities;
    \item In parallel with monetisation, a multicriteria analysis should be performed, recognising where quantities are not fully commensurate, i.e. they cannot naturally be brought together in the same line item of money;
    \item Visualisations such as a red-amber-green representation of the different options against the range of criteria may be very helpful in presentation to decision makers.  
\end{itemize}

\section{Conclusions}\label{sec:conc}

%\subsection{Other considerations}

%\cjd{I have not edited this section yet, beyond moving some text to the new discussion section. We need to decide what the key recommendations are!}

This paper has explored use of a range of risk model outputs as a basis for resource adequacy assessment and capacity procurement, including extensions to standard decision analysis pictures such as risk averse metrics and wider visualisations of risk profile. For the GB example considered, as the capacity of renewables connected to the system increases, not all changes to the risk profile  are  captured by expected value metrics such as LOLE and EEU. If a single summary statistic is required, then CVaR with respect to the distribution of energy unserved or loss of load duration is attractive due to the way CVaR generalises these standard expected value metrics -- however this may not reveal other aspects of risk profile such as how the same annual aggregate model output can be made up of a larger/smaller number of less/more severe days. 

Overall, these results provide a strong argument for finding ways to balance objectives in a transparent way that recognises the interests of decision makers, and we provide  examples of how this might be done, along with an extended discussion of their use in practical application -- a particular challenge  is the specification of transparent standards when it is not natural to construct a monovariate utility function. For the necessary combination of broad uncertainty management with decision analysis, a possible framework is a Bayes network-based decision support system as described in \cite{volodina}; any alternative would require similar functionality for combining a wide range of uncertainties.

\section*{Acknowledgments}

The authors acknowledge discussions with A. Dobbie, H. Wynn, S. Zachary, and members of the IEEE RAWG and the RA activities of ESIG, EPRI and G-PST. They also acknowledge meetings in which Colin Gibson and Andrew Wright provided advice about decision maker interests based on their industry experience. 

Author Chris Dent further expresses gratitude to the late Colin Gibson for  exchanges throughout CD's career in energy systems, from which he learned much  about  power system planning and operation. 

The authors acknowledge the following funding for this work: CJD, AS, JQS, ALW and XY from the Alan Turing Institute `Towards Turing 2.0' programme under  EPSRC Grant EP/W037211/1; CJD, JQS and ALW from the Turing Fellow project ‘Managing Uncertainty in Government Modelling’; NS a PhD scholarship from the Mexican Conacyt funding council. CJD, JQS and ALW also acknowledge Turing Fellowships from the Turing Institute; and CJD and JQS  would like to thank the Isaac Newton Institute for Mathematical Sciences for support and hospitality during the programme Mathematical and Statistical Foundations of Data Driven Engineering when work on this paper was undertaken (supported by EPSRC Grant Number EP/R014604/1), associated with which CJD  was partially supported by a grant from the Simons Foundation. CJD further acknowledges a grant from the International Centre for Mathematical Sciences under its Knowledge Exchange Catalyst scheme to work with the Global Power System Transformation consortium.

%\section*{References}

\bibliographystyle{SageV}
\bibliography{pesda.bib}

\end{document}